\documentclass[aps,pre,amsmath,amsfonts,amssymb,nofootinbib]{revtex4}

\usepackage{graphicx}
\usepackage{epsfig}
\usepackage{subfigure}
\usepackage{dcolumn}

\def\uS{\underline{S}}
\def\H{{\cal H}}

\begin{document}
\title{Finding long cycles in graphs}
\author{Enzo Marinari$^{\,1}$, Guilhem Semerjian$^{\,2}$ and Valery Van Kerrebroeck$^{\,1}$}
\affiliation{$^{1\,}$
Dipartimento di Fisica and INFN, Sapienza Universit\`a di Roma, 
P. A. Moro 2, 00185 Roma, Italy,
}
\affiliation{$^{2\,}$
LPTENS, Unit\'e Mixte de Recherche (UMR 8549) du CNRS et de l'ENS
associ\'ee \`a l'universit\'e Pierre et Marie Curie, 24 Rue Lhomond, 75231 
Paris Cedex 05, France.
}

\begin{abstract}
  We analyze the problem of discovering long cycles inside a graph. We
  propose and test two algorithms for this task. The first one is
  based on recent advances in statistical mechanics and relies on a
  message passing procedure. The second follows a more standard Monte
  Carlo Markov Chain strategy. Special attention is devoted to
  Hamiltonian cycles of (non-regular) random graphs of minimal
  connectivity equal to three.
\end{abstract}

\maketitle

\section{Introduction}

Random graph theory~\cite{Bollo,Janson,Wormald} is a fascinating branch of
mathematics, at the frontier between classical graph theory and probability.
Since the seminal work of Erd\"os and R\'enyi~\cite{ER} 
the properties of various random graph ensembles have been thoroughly studied.
Random graphs have been often encountered in the recent statistical mechanics
literature, in at least three contexts. In the first one, pioneered by 
Viana and Bray~\cite{VB}, they provide a convenient family of models of 
disordered systems which enjoy the remarkable analytical properties
of mean-field models, while retaining a limited notion of geometry lost
in the fully-connected models of the Sherrington-Kirkpatrick~\cite{SK} type.
A second family of statistical mechanics papers is also related to random 
graphs, but takes
a somehow inverse perspective: one studies carefully chosen disordered models
with the aim of obtaining a detailed description of the underlying random
graph ensemble (see for instance \cite{EnHaMo,MaMoSe,MaSe,MeZd,PrWe}). 
A third, yet related to the
second, family of works are the studies of real-world networks, where
random graphs are used as benchmarks for comparison to experimentally collected
data~\cite{review_networks}.

A large fraction of the above mentioned research has been concerned with
the presence of particular subgraphs (sometimes called patterns or motifs)
inside a given, possibly random, graph. For concreteness let us consider
the case where the pattern is a cycle of length $L$, i.e.~a closed, 
non-intersecting path of $L$ adjacent vertices. Several variants of the problem
can be stated. The first is the \emph{existence} problem.
For a given graph, for instance a real-world network, one can
ask whether or not there is such a cycle. In the random graph context, the
question amounts to compute the probability of presence of this pattern,
with respect to the choice of the graph in the random ensemble. In some
interesting cases this probability can present a threshold phenomenon; as
some external parameter defining the ensemble is varied, the limit of this
probability when the size of the graphs goes to infinity can be either zero
or one. A second, more demanding problem is the one of \emph{counting}: 
given a graph,
how many repetitions of the pattern can be found ? Or for a random graph 
ensemble, what is the distribution of the random variable counting this
number of occurrences ? A third version, which shall be the main focus of
this paper, is the \emph{finding} problem: can one
actually exhibit one example of the pattern in a proposed graph ?

The answer to these questions depends a lot on the length $L$ of the
cycles under consideration. As long as $L$ is much smaller than the 
total number $N$ of vertices, the various problems are relatively easy. 
Short cycles are not too numerous and can thus be counted by exhaustive 
enumeration~\cite{Johnson,Ginestra},
which provides as a side result an answer to the \emph{existence} and 
\emph{finding} problem.
Moreover for most random graph ensembles the distribution of the number of
cycles of finite length can be easily computed in the limit where the total
number of vertices diverges. These problems become much more difficult when 
one considers cycles whose length is a finite fraction of the
total number of vertices~\cite{ben,BiMa}, for instance Hamiltonian 
cycles, i.e.~those of length $N$. Indeed, such long cycles can be exponentially
numerous, preventing for instance the use of exhaustive enumeration when not 
dealing with very small graphs. 
More formally, deciding the existence of an 
Hamiltonian cycle is known to be an NP-complete problem~\cite{NP}. Because of 
the
large fluctuations of the exponential number of long cycles, their
distribution is known only in the special case of random regular 
graphs~\cite{Garmo} where all vertices have the same degree,
and which are known to be Hamiltonian with high probability~\cite{Ham_regular}.
In this restricted case there exists an algorithm for finding Hamiltonian 
cycles in polynomial time~\cite{counting_regular}.

The typical number of long cycles in a large class of random graphs has
been studied with statistical mechanics methods in~\cite{MaSe}. In
the present paper we extend this approach to the \emph{finding} version of
the problem, i.e.~we propose some algorithms which attempt to 
unveil large cycles in a graph. A special emphasis will be put on the
following issue. As mentioned above, random regular graphs of degree larger
than or equal to three are known to be Hamiltonian with high probability.
Wormald conjectured in~\cite{Wormald} that this remains true for random
graphs where all vertices have degrees in $[3,k_{\rm max}]$, with 
$k_{\rm max}\ge 3$ a finite integer. The non-rigorous approach of~\cite{MaSe}
reached the same conclusion and provided a prediction for the typical number 
of Hamiltonian cycles in graphs of these ensembles. In the present paper we
shall see that the algorithms we propose are indeed able to explicitly 
construct Hamiltonian cycles of these graphs.

The paper is organized as follows. Section~\ref{sec_def} is devoted to a
presentation of more formal definitions.
Two distinct approaches to the problem are presented in 
Section~\ref{sec_decimation} and \ref{sec_mc}, with technical details of the
implementation deferred to two appendices. We draw our conclusions in 
Section~\ref{sec_conclusions}.

\section{Definitions} 
\label{sec_def}

An undirected graph $G$ is defined by a set $V$ of $N$ vertices, and a set 
$E$ of 
$M$ unordered pairs of vertices, the edges. 
The graphs considered in this article are
simple, i.e.~neither self-loops (edges from one vertex to itself)
nor multiple edges between the same pair of vertices are allowed.
Two vertices $i$ and $j$ are said to be adjacent if the edge $\{i,j\}$ belongs
to $E$. The degree (or connectivity) of a vertex is the number of edges
it belongs to. The neighborhood $\partial i$ of vertex $i$ denotes
the set of edges in which $i$ appears, or the set of adjacent vertices,
without possibility of confusion. 

A cycle (or circuit) of length $L$ is a closed non-intersecting path of $G$,
more formally a sequence of $L$ distinct vertices $(i_1,\dots,i_L)$ such
that $i_n$ is adjacent to $i_{n+1}$ for all $n \in [1,L-1]$, and $i_1$ is 
adjacent to $i_L$. A graph $G$ is said to be Hamiltonian if it admits 
a Hamiltonian cycle, i.e.~a cycle of length $N$ which visits
all the vertices of $G$.

We identify a subgraph of $G$, a cycle for instance, 
by the set of its edges. For the
ease of notation, let us associate to each edge $\{i,j\}=l \in [1,M]$ of
$G$ a discrete variable $S_l\in \{0,1\}$. Each of the $2^M$ possible 
subgraphs of $G$ is
unambiguously associated to a configuration $\uS=\{S_1,\ldots,S_M\}$,
with $S_l=1$ (respectively 0), if edge $l$ is present (absent) in the
subgraph. We also introduce the notation 
$\uS_i=\{S_l|l \in \partial i\}$ for the configuration of the edges around
vertex $i$.

Besides cycles as defined above, we shall also encounter subgraphs of $G$
which are unions of vertex disjoint cycles. When all the vertices of the
graph are covered in this way, the subgraph is called a cycle cover, 
or a 2-factor. By definition Hamiltonian cycles are cycle covers 
made of a single cycle, but obviously not all cycle covers are Hamiltonian 
cycles. Even if apparently similar, the decision problems concerning 
the existence of a Hamiltonian cycle or of a cycle cover have very
different computational complexity. The first one is in the NP-complete 
class~\cite{NP},
whereas the second can be mapped, thanks to a theorem of 
Tutte~\cite{Tutte}, to the existence of 
a perfect matching in a dual graph of similar size, a task for which 
polynomial time algorithms are known~\cite{matchings}.

We shall test our algorithms on graphs drawn from the fixed 
degree distribution ensemble, discussed for instance in~\cite{MR,NSW}. 
We choose our random graphs uniformly 
among those with degree distribution $q(k)$, a given probability 
distribution on positive integers $k \ge 3$ \footnote{Vertices of degree 0 and
1 can indeed be eliminated by reducing the graph to its 2-core, outside of 
which no cycles can be drawn. When a finite fraction of degree 2 vertices
is allowed, the graphs are expected to be non Hamiltonian with high 
probability~\cite{MaSe}. The longest cycles of such graphs could also 
be studied, at
the price of some technicalities we decided to avoid.}.
The generation of such graphs, based on the configuration 
model~\cite{config}, is done as follows. For each value of $k$ a subset
of $N_k=N q(k)$ vertices are assigned degree $k$, and $k$ ``half-links'' are
drawn around each of them. 
The half-links are then paired in a uniform random way. 
If the graph generated is not simple, it is discarded and the generation
starts again. It can be shown~\cite{Wormald} that this procedure respects
the uniformity over the graphs with prescribed degree distribution. 
Regular random graphs of degree $c$ are a particular case of the above model,
where all vertices have the same degree $c$, i.e.~$q_c(k)=\delta_{k,c}$. We
shall also consider the distribution 
\begin{equation}
q_{c_1,c_2}^{\epsilon}(k) = 
(1-\epsilon)\;\delta_{k,c_1} + \epsilon \;\delta_{k,c_2},
\label{eq:degree_distrib_mixture}
\end{equation}
 which 
interpolates between two different regular
ensembles as $\epsilon$ varies in $[0,1]$.

The goal of the algorithms presented in the following is to discover the
longest cycles in a given graph. The class of graphs we investigate
(random with minimal degree larger than 3 and bounded maximal degree) are
expected~\cite{Wormald,MaSe} to be Hamiltonian with high probability when 
their size diverges. We thus aim at finding Hamiltonian cycles, or at least
cycles covering almost all vertices of the graphs. Because of the intrinsic
complexity of the Hamiltonian circuit problem we do not expect that these
algorithms should be valid for all graphs, but at least for the class of
sparse graphs satisfying the degree constraints explained above.

\section{Belief Inspired Decimation Algorithm}
\label{sec_decimation}

\subsection{Description of the Algorithm}
\label{sec:bid_description}
Let us first introduce in generic terms the principles underlying the 
algorithm we develop in this section. Consider an arbitrary discrete set of 
configurations $\uS=\{S_1,\dots,S_M\}$, a subset $\H$ that one would
like to sample and define ${\rm Prob}[\uS]$ as the uniform probability
measure on $\H$, i.e.~${\rm Prob}[\uS] = 1 /|\H|$ if $\uS \in \H$ and zero
otherwise, where $|\H|$ is the number of configurations in $\H$. A possible
scheme for the sampling from $\H$ is the following. 
Initially all $S_l$ are undetermined. For $n$ increasing from
1 to $M$, choose arbitrarily an index $l_n$ of one of the still non fixed
variables, and draw $S_{l_n}$ according to its marginal law conditioned on 
the previous choices, ${\rm Prob}[S_{l_n} | S_{l_1},\dots,S_{l_{n-1}}]$ 
(for $n=1$ there is no conditioning). The configuration produced at the
end of this ``decimation'' process is clearly uniformly distributed on 
$\H$. However, except in particularly simple situations, it is not possible
to implement this method as it is, the marginal laws used above being in 
general very difficult to compute exactly.

In the context of this paper we would ideally like to follow this road taking
for the set $\H$ the longest cycles of the graph under study. The output
configuration would then provide us with the length of these longest cycles and
one of their representatives. There is no serious hope for a practical
implementation of this idea in an exact way: note for instance that it would,
as a side result, solve the Hamiltonianicity decision problem, which, being
NP-complete, is not expected to have a polynomial time algorithm.

Therefore we turn to an approximated version of this ideal strategy,
similar in spirit to the Survey Propagation algorithm introduced by
M\'ezard and Zecchina for Constraint Satisfaction Problems~\cite{SP}.
We first define a probability law on the subgraphs of $G$,
\begin{equation}
{\rm Prob}[\uS] = \frac{1}{Z} u^{\sum_l S_l} \prod_{i} w_i(\uS_i) \ ,
\label{eq:prob}
\end{equation}
where $w_i (\uS_i) = 1$ if $\sum_{l \in \partial i}S_l \in \{0,2\}$, 
$w_i(\uS_i)=0$ otherwise, $u$ is a real parameter and $Z$ a normalization
constant.

This allows only subgraphs made of vertex disjoint cycles (each node of the
graph must be surrounded by either 0 or 2 edges). Their weight is proportional
to $u^L$, where $L$ is the total length of these cycles. In case $u=1$ this 
leads to a flat sampling amongst vertex disjoint cycles of any length. When $u$ goes to
$+ \infty$, the probability converges to the uniform law on the largest of 
these subgraphs, in particular cycle covers if the graph admits them.
Allowing for unions of several cycles leads us away from the ideal strategy but
is a gain in analytical simplicity: the interactions in Eq.~(\ref{eq:prob})
are local, i.e.~involve only a finite set of neighboring
variables. This opens the way to several approximate treatments, of the 
mean-field Bethe approximation flavor. We shall indeed use 
Belief Propagation (BP), a message passing algorithm widely used for
solving inference problems (see~\cite{Yedidia} for a review and for the 
connection with the Bethe free-energy). This approach was followed 
in~\cite{MaSe} in order to compute the normalization constant $Z$, and
hence the number of allowed configurations. One can easily adapt these
results for the relevant question here, i.e.~the computation of the marginal
probabilities of presence of an edge $l$ in the law (\ref{eq:prob}) 
conditioned on some of the variables being fixed. 

When using the BP algorithm, we work directly in the limit $u \to +\infty$, 
where the measure concentrates on the longest configurations. In this limit, 
the fact that sites of degree 2 are not allowed implies a number of 
analytical simplifications~\cite{MaSe}.
Let us consider the set $E$ of edges of the graph. We assume that some of them 
are constrained to be present, i.e.~have $S_l=1$, and call the set of these edges $E_1$, 
and that some of them are constrained to be absent, i.e.~have $S_l=0$, and call this set $E_0$.
The edges of
$E_* = E\setminus (E_1 \cup E_0)$ are called non-decimated. 
We introduce for each non-decimated edge $l=\{i,j\}$ a pair of real variables 
$y_{i \to j}$ and $y_{j \to i}$, ``messages'' sent by vertex $i$ to vertex $j$ 
and vice versa. The BP estimate for the probability of presence of edge
$l$ in this conditional law reads
\begin{equation}
p_l = {\rm Prob}[S_{l}=1 | E_0, E_1] =  
\frac{y_{i \rightarrow j}y_{j \rightarrow i}}
{ 1+ y_{i \rightarrow j}y_{j \rightarrow i}} ,
\label{eq:marginal}
\end{equation}
where the messages are solutions of the BP equations. These express the value
of a message $y_{i \to j}$ in terms of the messages along the neighboring edges and
the status (decimated or not) of the edges. Let us denote 
$\partial_* i \setminus j$ the set of vertices $k$ adjacent to $i$, distinct 
from $j$ such that $\{i,k\}$ is a non-decimated edge. We have to consider two cases:
\begin{itemize}
\item[$\bullet$] if all the edges in $\partial i \setminus j$ are either 
non-decimated or constrained to be absent,
\begin{equation}
y_{i \to j} = \frac{
\underset{k \in \partial_* i \setminus j}{\sum} y_{k \rightarrow i}}
{\frac{1}{2} 
\underset{k\neq k'}{\underset{k,k' \in \partial_* i \setminus j}{\sum}}
y_{k \to i}y_{k' \to i}}
\label{eq:msg1}
\end{equation}
\item[$\bullet$] if exactly one edge in $\partial i \setminus j$ is 
constrained to be present, the others being either non-decimated or absent,
\begin{equation}
y_{i \to j} =
\frac{1}{\sum_{k \in \partial_* i \setminus j} y_{k \to i}}
\label{eq:msg2}
\end{equation}
\end{itemize}
If two edges in the neighborhood of $i$ are constrained to be present, the others are set to absent and the messages along these edges are no longer considered in the BP algorithm. Cases in which three or more edges incident to $i$ are present are not allowed due to the constraints $w_i(\uS_i)$.    
We refer to \cite{MaSe} for details on the derivation
of these equations.

We can now describe a possible implementation of the proposed algorithm. 
Initially, all edges are  non-decimated ($E_0=E_1=\emptyset$)
and we set all messages to a random value in $[0,1]$.
 The algorithm then proceeds by repeating the following two
steps:
\begin{itemize}
\item[$\bullet$] Belief Propagation step.

A solution of the BP equations is searched by iterating 
Eqs.~(\ref{eq:msg1},\ref{eq:msg2}); all messages are updated in random 
sequential order. This operation is repeated a certain number of times to
get sufficiently close to the fixed-point solution. In our implementation
we stopped the iterations when either the average modification of a message
is lower than an arbitrary small threshold (set to $10^{-6}$ in the following)
or when a maximum number (20) of iterations fixed beforehand has been reached.

\item[$\bullet$] Decimation and propagation step.

We set some of the non-decimated edges to either present or absent
(updating accordingly $E_0$ and $E_1$), according to the information given by 
the solution of the BP equations found in the previous step.  

If the BP procedure were exact, one could choose arbitrarily one of the
edges in $E_*$ and fix it according to its marginal probability (conditional
on $E_0,E_1$). However the BP estimation of this quantity $p_l$
(see Eq.(\ref{eq:marginal})) is only approximate.
Therefore it is safer to fix
only the most biased variable, i.e.~the one with $p_l$ closer to 0 or
1, to its most probable value, hoping that this is the least subject
to the imprecision of the algorithm.

Because of the hard constraints encoded in the weight function $w_i$,
fixing a variable might automatically impose the value of a few others
(similarly to the Unit Propagation rule in Constraint Satisfaction
Problems).  For example, when two edges around a given vertex have
been assigned to present, the other neighboring edges have to be
absent in order to avoid intersecting cycles. Similarly,
if all but two edges around a given vertex are absent, we set these
two non-decimated edges to be present since we are seeking configurations
of maximal length. In case we encounter a situation where only one
edge is left undecimated, we set it to present if there is already
another neighboring edge present.  Otherwise, we set it to absent,
such that the final subgraph could still be a cycle (even though not a
Hamiltonian one). It can happen that this propagation leads to a
contradictory situation in which an edge needs to be both present and
absent. When this occurs, the decimation procedure is stopped
prematurely, we re-initialize all messages and all edge variables and
re-start the complete decimation procedure from the beginning.

Apart from the effects of this direct propagation, decimating a single
variable does not drastically modify the values of the messages on the
remaining non-decimated edges. Thus, 
it turns out to be useful to fix several of the most biased variables before
returning to the BP step with a reduced graph. In practice, we treat all the
edges with $p_l$ lower than $0.2$ or higher than $0.9$ in a single decimation
step and if there are no such edges, we simply choose one of the most biased ones.
\end{itemize}
We repeat these two steps subsequently until
either a contradiction has occurred during a propagation step, 
or all edge variables have been assigned a value. 
We refer to the complete operation as the decimation procedure.
If the procedure is exited without contradiction, we are left with a 
subgraph made of vertex disjoint cycles, 
which can be a cycle cover or possibly a Hamiltonian cycle. 

The output of a decimation procedure is stochastic, since we use 
random numbers in the initialization of the BP messages and in deciding 
the order of the updates. The algorithm can thus be improved in a 
straightforward way. If the decimation procedure result is not satisfactory, 
one can launch it again with different random numbers, until one of these 
decimation procedures produces a satisfactory output (cycle cover or 
Hamiltonian cycle). We set a maximum number of repetitions equal to $1000$.
The influence of this arbitrarily chosen number is discussed in the following 
section.

We shall also introduce another improvement based on a local 
rewiring procedure, that will be interleaved between the repetitions of the 
decimation procedures, see below for details.

Before entering the discussion of the results of the algorithm, let us
note that, in the family of graphs studied here, one decimation
procedure has a computational cost at most quadratic in the size of
the graph. Since for the sparse graphs we are considering the number
of edges is proportional to the number of vertices, a BP step has
linear cost. In a worst case scenario, only one edge is fixed during
each subsequent decimation step, resulting in at most $M$ repetitions
of the BP step.

\subsection{Numerical Results}
\label{sec:success}

\begin{table*}
\begin{ruledtabular}
\begin{tabular}{c|c|c|c|ccc|ccc|ccc}
 &$q_3$&$q_4$&$q_5$&\multicolumn{3}{c|}{$q_{3,4}^{0.5}$}&\multicolumn{3}{c|}{$q_{3,5}^{0.5}$}&\multicolumn{3}{c}{$q_{4,5}^{0.5}$}\\
 &HC&HC&HC&CC&\multicolumn{2}{c|}{HC}&CC&\multicolumn{2}{c|}{HC}&CC&\multicolumn{2}{c}{HC}\\
 N& & & & & DEC & LR & & DEC & LR & & DEC & LR \\ \hline
 100 & 100.0 & 100.0 & 100.0 & 99.9 & 96.0 & 99.6 & 98.9 & 69.9 & 92.9 & 98.7 & 56.9 & 96.0 \\ 
 200 & 100.0 & 100.0 & 100.0 & 99.6 & 96.2 & 99.3 & 99.7 & 71.1 & 95.2 & 98.9 & 50.0 & 96.0 \\ 
 400 & 100.0 & 100.0 & 100.0 & 99.7 & 96.4 & 99.2 & 99.9 & 67.7 & 95.4 & 98.9 & 50.7 & 94.2 \\ 
 800 & 100.0 & 100.0 & 100.0 & 99.8 & 96.7 & 98.7 & 99.6 & 68.9 & 95.7 & 99.6 & 46.8 & 94.5 \\ 
 1600& 100.0 & 100.0 & 100.0 & 99.7 & 97.8 & 98.7 & 99.9 & 68.6 & 92.0 & 99.9 & 52.3 & 94.0 \\ 
\end{tabular}
\end{ruledtabular}
\caption{\label{tab:prob_cc_hc} Percentage of cycle covers (CC) and Hamiltonian cycles (HC) found with the decimation procedure (DEC), possibly combined with the local rewiring procedure (LR), for various connectivity distributions ($q_c(k)=\delta_{k,c}$ or is defined as in Eq. 
(\ref{eq:degree_distrib_mixture})) and graph sizes. Each entry was obtained by investigating a thousand different samples.}
\end{table*}

We present in Table~\ref{tab:prob_cc_hc} a summary of the numerical
experiments we conducted with the decimation algorithm. Each entry of
the table is a percentage of success computed on a set of one thousand
graphs of given size and connectivity distribution.  In the
CC column we define successful a repetition of decimation procedures
 that terminates with the finding of a Cycle Cover (CC),
while in the HC column we define by success the discovery of an
Hamiltonian Cycle (HC).
 
Consider first the three leftmost columns of Table
\ref{tab:prob_cc_hc}, concerning regular graphs of degree 3, 4 and
5. For these cases we employed the decimation procedures in the
simplest way: on each graph we repeated a decimation procedure until
the output configuration was a Hamiltonian cycle, or until a maximal
number of trials had been reached. This approach turns out to be very
efficient. For all the regular graphs investigated, we find an
Hamiltonian cycle after about 10 trials on average, i.e.~well before
reaching the cutoff of a thousand repetitions.  The efficiency of this
procedure is a priori surprising. If the marginal probabilities were
computed exactly the output configurations would be distributed among
the various cycle covers of the graph, a set which contains the
Hamiltonian cycles but could be much larger. It was however argued
in~\cite{MaSe} that, at the leading exponential order, these two sets
have roughly the same size (rigorous arguments of this kind were used
in~\cite{counting_regular,fpras_dense}). Hence, by generating enough
distinct cycle covers one of them is eventually a Hamiltonian cycle.

This simple procedure is less efficient on the non-regular graphs
(with a degree distribution given by
Eq. (\ref{eq:degree_distrib_mixture})) that we investigated, as shown
in the three rightmost groups of columns of
Table~\ref{tab:prob_cc_hc}.  The percentage of graphs for which an
Hamiltonian cycle is found within the first thousand repetitions of
the decimation procedure is shown in the columns HC-DEC. When the
graphs get denser the success probability drops.

A more detailed look at the outputs of the algorithm
reveals that in almost all of the graphs a cycle cover has however
been found (columns CC). This motivated the search for a simple way
to convert a cycle cover, obtained with the decimation procedure
and composed of several vertex disjoint cycles, into a Hamiltonian
cycle. The main idea is to join two or more cycles by doing some local
rewiring (LR) among the edges: we discuss the details of such rewiring
in the appendix~\ref{app:LR}.  In practice, when the decimation
procedure ends with a cycle cover different from a Hamiltonian cycle,
we apply this LR algorithm to try to recover a Hamiltonian
Cycle.  Again, if this extended decimation procedure does not
produce the desired result, it is repeated no more than a 
thousand times.

The success rate of finding Hamiltonian cycles with this combined
strategy (columns HC-LR) is much higher with respect to the simple one 
(HC-DEC). In fact, by using this extended version, it now almost
matches, up to a few percentage points, the fraction of graphs for
which we were able to find a cycle cover.  The added value of
including this local rewiring algorithm reduces slightly with the size
of the graph and when the average graph degree increases.  Note however
that even in ``half-successful'' graphs, i.e. when a cycle cover but no
Hamiltonian cycle is found, the cycle cover always contains one 
long (extensive) cycle, whose length gets closer to $N$
for increasing graph sizes.

\medskip

Let us now evaluate the efficiency of the algorithm, which necessarily
turns out to be a trade-off between the success rate and its time
requirements.  We have to distinguish two points.

\begin{figure}[!ht]
  \centering
  \includegraphics[angle=270]{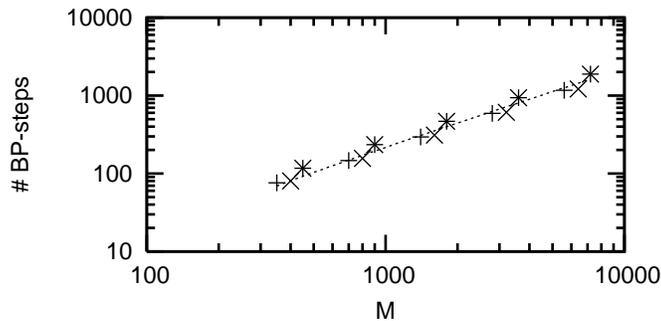}
  \caption{\label{fig:nb_MP}  Average number of steps needed for the
    decimation procedure to exit for graphs of different
    connectivity distributions 
    ($+$ for $q_{3,4}^{0.5}$, $\times$ for $q_{3,5}^{0.5}$ and
    $*$ for $q_{4,5}^{0.5}$) and sizes. The best fit to all data
    for all connectivity distributions is linear and has a slope 
    of $0.23$.}
\end{figure}

A first point was already mentioned at the end of
Sec.~\ref{sec:bid_description} and concerns the computational cost of
a single decimation procedure, that we argued to be quadratic in the
size of the graph. This is confirmed by the plot of Fig.~\ref{fig:nb_MP}:
the average number of decimation steps performed during a decimation
procedure scales linearly. Moreover, as we allowed several variables to
be fixed during a single decimation step, this number is smaller than
$M$, the worst-case estimate.

A second crucial point concerns our choice (one thousand)
for the threshold 
on the number of repetitions
of the complete decimation procedure (that can be
possibly complemented by an attempt of
patching the cycle cover). This is the moment when
we give up our search of a Hamiltonian
cycle: obviously such a choice 
has a direct effect on the percentages of success 
we presented in Table \ref{tab:prob_cc_hc}. 

\begin{figure}[!ht]
  \centering
  \subfigure{\includegraphics[angle=270]{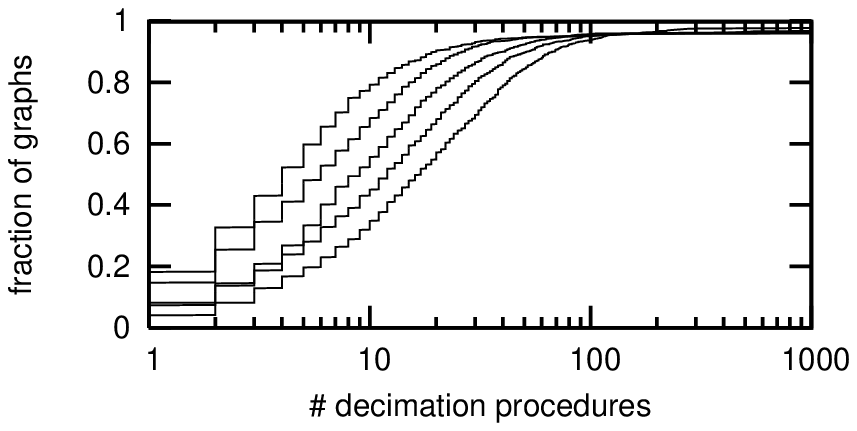}}
  \subfigure{\includegraphics[angle=270]{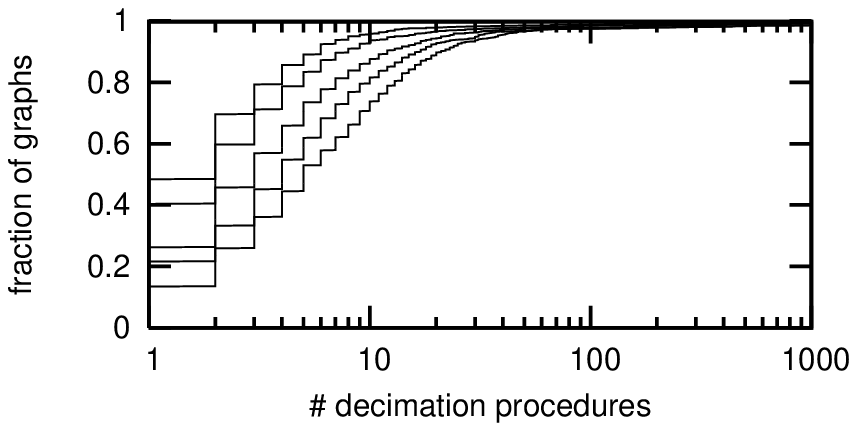}}
  \caption{\label{fig:nb_trials} Integrated distribution of the number
of decimation procedures needed to find a Hamiltonian cycle for
graphs with connectivity distribution $q_{3,4}^{0.5}$
of various sizes ($N$ increases from left to right). 
Left/Right : with/without the use of the patching procedure.
}
\end{figure}

\begin{figure}[!ht]
  \centering
  \subfigure{\includegraphics[angle=270]{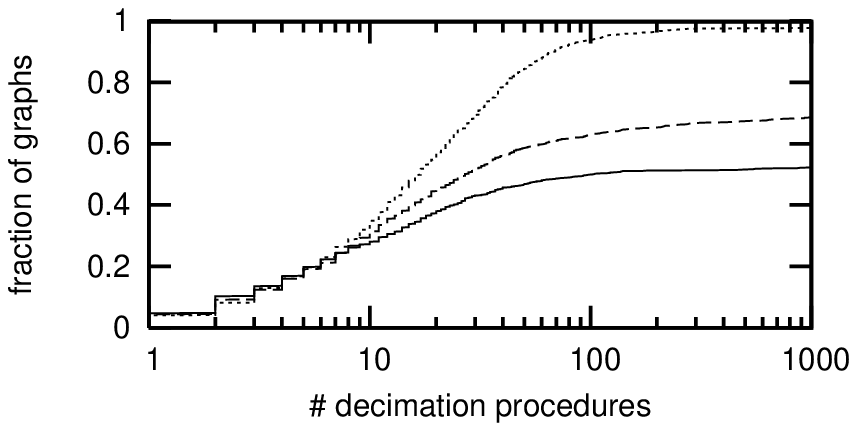}}
  \subfigure{\includegraphics[angle=270]{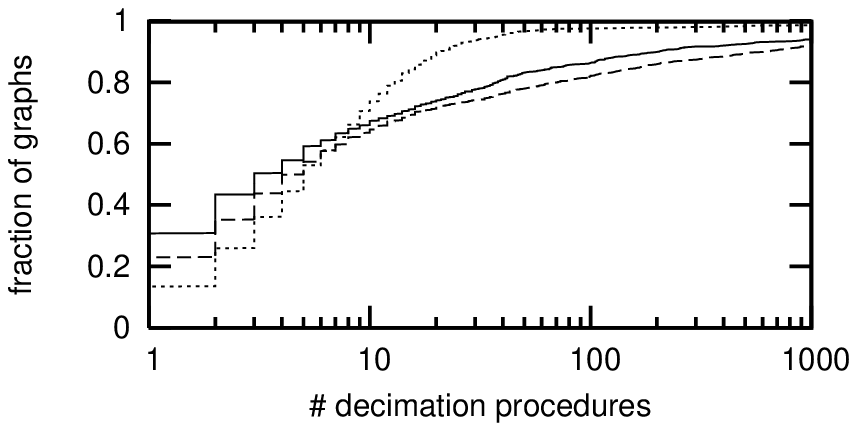}}
  \caption{\label{fig:nb_trials_varyq} Integrated distribution of the number
of decimation procedures needed to find a Hamiltonian cycle for
graphs with connectivity distribution $q_{3,4}^{0.5}$ (dotted lines),
$q_{3,5}^{0.5}$ (dashed lines) and
$q_{4,5}^{0.5}$ (full lines) of sizes $N=1600$.
Left/Right : with/without the use of the patching procedure.}
\end{figure}

In Fig.~\ref{fig:nb_trials} and \ref{fig:nb_trials_varyq}  
we plot the integrated distributions of
the number of decimation procedures (alone or combined with the
patching algorithm) performed before exiting.  These
distributions are artificially bounded by the threshold we set, i.e.~by
the maximum of thousand repetitions of the decimation procedure.  From
these figures one can learn, for example, how the success probability
would deteriorate by taking a smaller threshold on the number of
repetitions. Consider first the right panel of Fig.~\ref{fig:nb_trials},
displaying the results of the combined strategy for graphs of connectivity
distribution $q_{3,4}^{0.5}$. It is clear that the almost constant plateau
of success probability is already reached around one hundred repetitions.
Lowering the threshold on the number of repetitions to this value would
reduce the success probability stated in Table \ref{tab:prob_cc_hc}
by roughly one percent only (more precisely $1.2$ percent for $N=1600$
and $0.7$ percent for $N=100$), justifying ``a posteriori'' our choice.

Let us now compare the results obtained by using only the decimation
strategy to the ones where we also integrated the patching procedure,
 i.e.~the left and
right part of Fig.~\ref{fig:nb_trials} respectively. For this connectivity
distribution the difference in the success probability (the fraction of
solved graphs for the threshold of 1000 repetitions) is not drastic. 
However, the introduction of the patching does reduce the median number 
of repetitions before finding a Hamiltonian
cycle roughly by a factor three, as can be seen from the shift 
of the distributions from one panel to the other. Indeed, the use of the intermediate local rewiring step can only enhance the probability of success of
a decimation procedure.

The dependence of these results on the connectivity distribution is
illustrated in Fig.~\ref{fig:nb_trials_varyq}. Two points are worth signaling
for the densest graphs. The success probability (for a maximum repetition of 
1000) is largely enhanced by the local rewiring procedure (as was already
mentioned in the discussion of Table \ref{tab:prob_cc_hc}). Also, its
deterioration by reducing the repetition threshold is more important
than for the graphs of connectivity distribution $q_{3,4}^{0.5}$. Indeed
the plateau in the right panel of Fig.~\ref{fig:nb_trials_varyq} is
less flat, meaning that in this case augmenting the threshold should still
improve slightly the success probability.

It is not easy to assert how the efficiency of the method evolves with the
size of the graphs. The median of the 
distributions of the number of needed repetitions (see for instance 
Fig.~\ref{fig:nb_trials}) can be fitted, on the interval of $N$ we studied,
by a power of $N$ with an exponent smaller than one (different for the
various connectivity distributions studied). 
This leads to a rough ``less than cubic'' estimation for the computational 
cost of the (repeated) decimation strategy.
For concreteness we mention that for
a typical random graph of size $N=1600$ and degree distribution
$q_{3,4}^{0.5}$ our implementation takes about $30$ minutes 
to locate a Hamiltonian cycle by using the decimation strategy, 
which narrows down to a mere $5$ minutes when local rewiring is included 
(here and in the
following the execution times are given for a computer with a $2.00$ GHz 
Intel Pentium M processor and $1.5$Gbyte of RAM).

\section{Markov Chain Monte Carlo Algorithm}
\label{sec_mc}

\subsection{Description of the Algorithm}

The second approach we describe in this section follows an idea
largely used and studied in statistical mechanics and computer
science. If one wants to sample from a given probability measure ${\rm
Prob}[\uS]$, one can construct a Markov chain which admits ${\rm
Prob}[\uS]$ as a unique stationary distribution (for instance by
imposing detailed balance conditions on the transition probabilities). Several
issues have to be addressed for this simple idea to be turned into a
practical algorithm. One is the problem of ergodicity: the allowed
transition rates must prevent the chain from being stuck in some parts
of the configuration space apart from the interesting one which bears
the dominant contribution to the stationary measure. 
A second problem is the convergence time of the chain, which should be
small for the stationary measure to be reached in a
reasonable time. A large amount of research in theoretical computer
science has been devoted to this question, with a formal definition of
the mixing time of the chain and various methods for bounding
it~\cite{review_mcmc}; many results have also been obtained from the
side of statistical mechanics~\cite{review_mcsm}.  Finally a
compromise must be found between the simplicity of the allowed moves
in configuration space and their efficiency to explore it. Most
algorithms based on this idea are local, i.e.~the current
configuration of the chain is modified in a single variable (or in a
finite number of variables), with rates depending only on the status
of nearby other variables. As a notable exception we mention the
cluster algorithms~\cite{cluster_mc}, which are however restricted to
particular cases. It is also important to quote improvements to the
standard Monte Carlo approach like Tempering and Parallel
Tempering~\cite{pt} that allow numerical simulations of systems with a
complex phase space (with and without quenched disorder).

The authors of~\cite{ben,mc_circuits} have presented
a Monte Carlo (MC) simulation
method in the context of cycles in graphs. The approach
presented here is however different in the goal (we concentrate on the
finding instead of the counting problem), and in the means.
We shall use simple moves and consider a stochastic process 
in the space of subgraphs where each transition consists of the addition
or the removal of a single edge. As the initial and final configurations
in such a step can never be both unions of vertex disjoint
cycles, we have to relax the probability measure used before
(cf. Eq.~(\ref{eq:prob})).  We introduce instead:
\begin{equation}
{\rm Prob}[\uS] = \frac{1}{Z} u^{\sum_l S_l} \ \eta^{n_{\uS}} \ 
\prod_{i} \widetilde{w}_i(\uS_i)\;, 
\label{eq:relaxed_prob}
\end{equation}
where $n_{\uS}$ is the number of disjoint components in the configuration 
$\uS$, $\eta \in [0,1)$ is an external parameter, and the vertex weight 
$\widetilde{w}_i$ now allows for open paths,
\begin{equation}
\widetilde{w_i}(\uS_i) = \delta\left(\sum_{l \in \partial i} S_l\right) + \delta\left(\sum_{l \in \partial i} S_l - 2\right) + \epsilon \delta\left(\sum_{l  \in \partial i} S_l - 1\right).
\label{eq:relaxed_vertex_weight}
\end{equation}
This probability law
depends on 3 parameters: $u$, $\epsilon$ and $\eta$. 
A valid configuration is characterized by its three conjugate
observables, i.e.~the total length $L=\sum_l S_l$ of the subgraph $\uS$, 
the total number of vertices with exactly one occupied neighboring edge 
$n_{\epsilon}$ and the number $n_{\uS}$ of disjoint components of $\uS$, 
respectively. Eq.~(\ref{eq:prob}) is a special case 
of Eq.~(\ref{eq:relaxed_prob}) with $\epsilon = 0$ and $\eta = 1$. 
When $u \to + \infty$, $\epsilon \to 0$ and $\eta \rightarrow 0$, 
the law (\ref{eq:relaxed_prob}) concentrates on the longest single cycles 
of the graph, rather than the longest union of cycles. 

\begin{figure}[!ht]
  \centering
  \subfigure{\includegraphics[width=0.1 \textwidth]{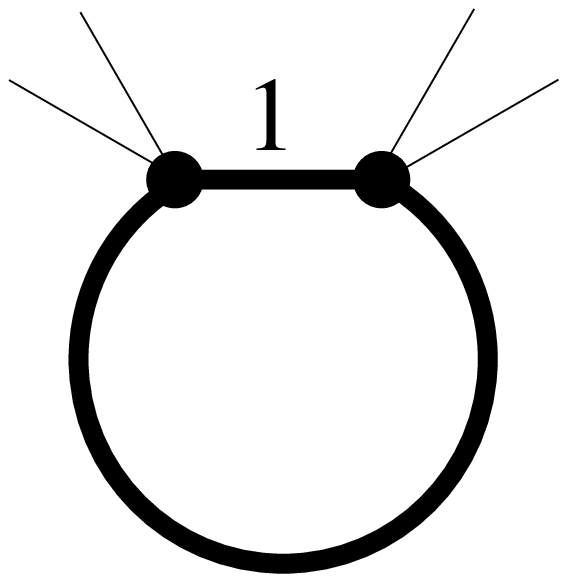}}
  \subfigure{\includegraphics[width=0.1 \textwidth]{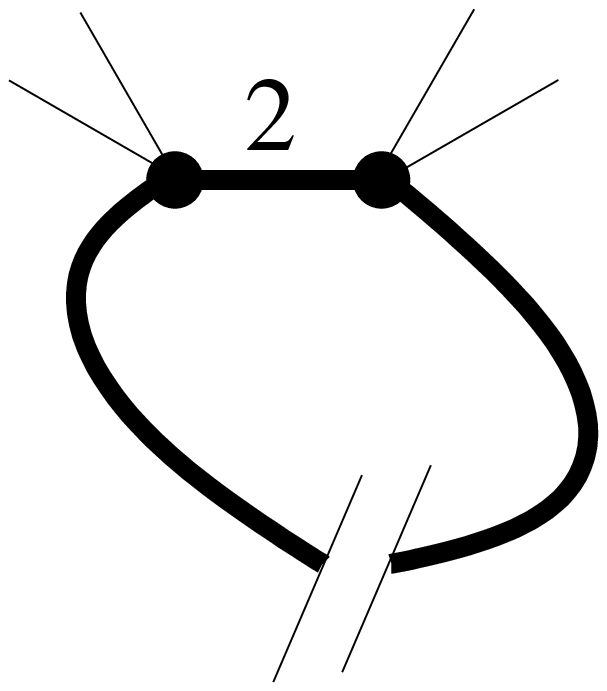}}
  \subfigure{\includegraphics[width=0.1 \textwidth]{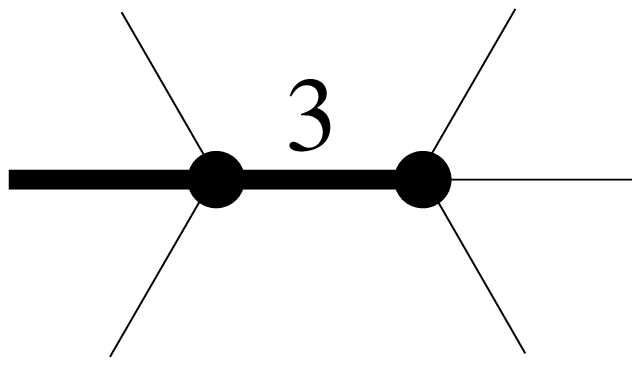}}
  \subfigure{\includegraphics[width=0.1 \textwidth]{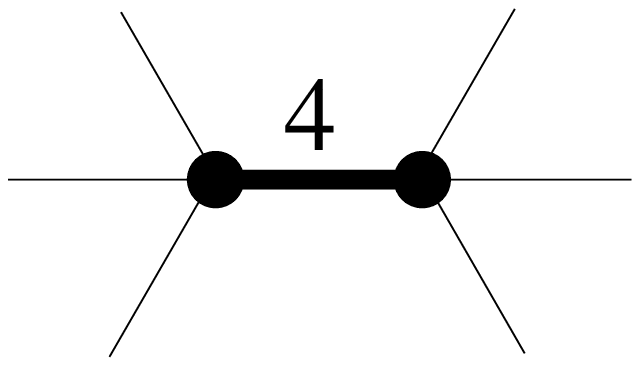}}
  \subfigure{\includegraphics[width=0.1 \textwidth]{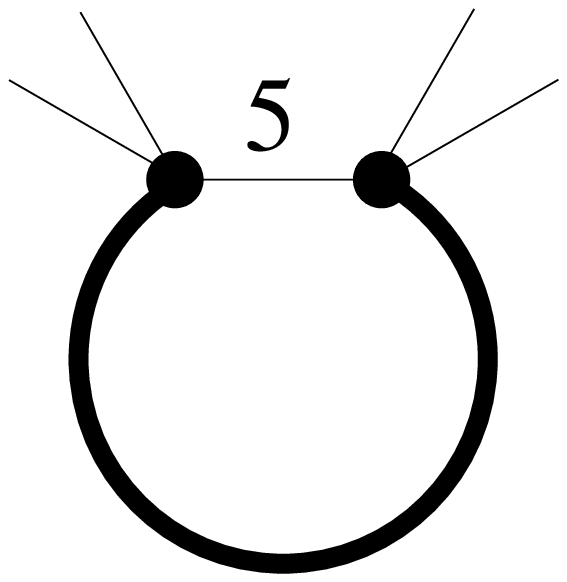}}
  \subfigure{\includegraphics[width=0.1 \textwidth]{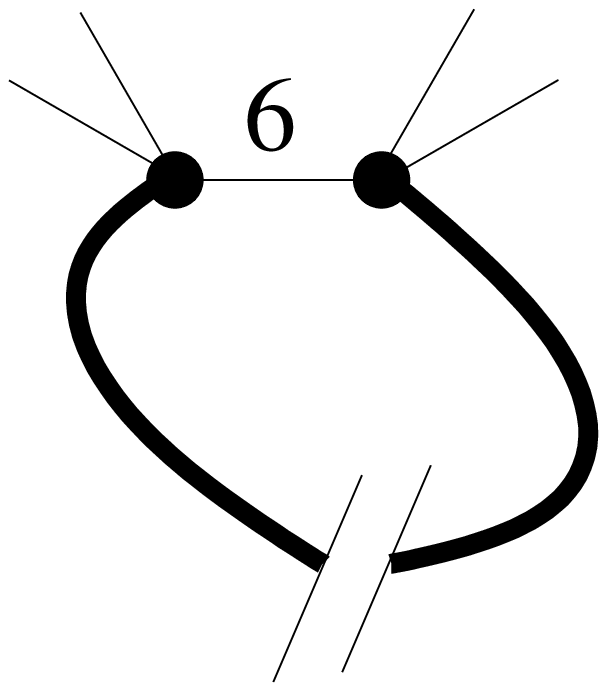}}
  \subfigure{\includegraphics[width=0.1 \textwidth]{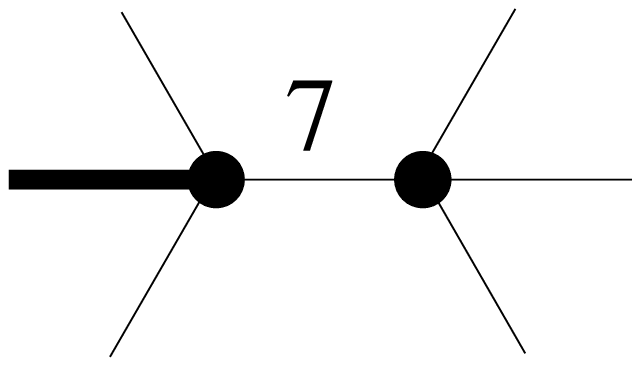}}
  \subfigure{\includegraphics[width=0.1 \textwidth]{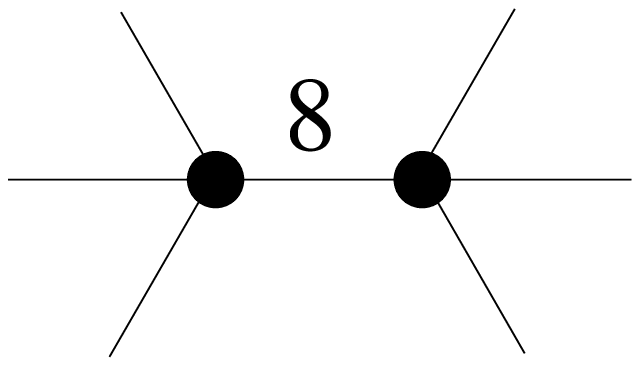}}
  \subfigure{\includegraphics[width=0.1 \textwidth]{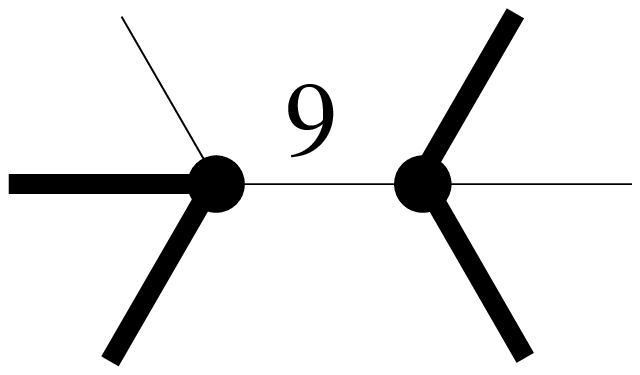}}
  $w_{1 \rightarrow 5}=\frac{\epsilon^2}{u}$\;\;\;\;
  $w_{2 \rightarrow 6}=\frac{\epsilon^2\eta}{u}$\;\;
  $w_{3 \rightarrow 7}=\frac{1}{u}$\;\;\;
  $w_{4 \rightarrow 8}=\frac{1}{u\epsilon^2\eta}$\;\;
  $w_{5 \rightarrow 1}=1$\;\;\;\;\;
  $w_{6 \rightarrow 2}=1$\;\;\;\;
  $w_{7 \rightarrow 3}=1$\;\;\;\;
  $w_{8 \rightarrow 4}=1$\;\;\;\;\;\;\;\;\;\;\;\;\;\;\;\;\;\;\;\;\;\;\;\;
  \caption{\label{fig:edgestate} Possible states $j$ of the central edge with all possible corresponding non-zero transition probabilities $w_{j \rightarrow i}$. The bold lines stand for an edge, path or cycle which is present, the thin lines mean it is absent. Two thin parallel lines separating present edges or paths signify those edges or paths are disconnected.}
\end{figure}

We now construct a MC algorithm that admits Eq.~(\ref{eq:relaxed_prob}),
for finite values of $u$, $\epsilon$ and $\eta$, as a stationary measure. 
More precisely, a
Monte Carlo sweep consists of $M$ steps, where in each step an edge index 
$l$ is drawn at random among the $M$ possible ones. Denoting $\uS$ the current 
configuration, a possible transition to the configuration $\uS'$ in which 
the status of the edge variable $S_l$ is reversed (from present to absent
or vice versa) is proposed and accepted with probability 
$W[\uS \to \uS']$. We impose the detailed balance condition on these
transition probabilities,
\begin{equation}
W[\uS \to \uS']{\rm Prob}[\uS] = W(\uS' \to \uS){\rm Prob}[\uS'] \ .
\label{eq:detailed_balance}
\end{equation}
The fact that the vertex weight $\widetilde{w_i}$ strictly allows only three 
different vertex neighborhoods reduces the number of possible transitions 
drastically. We illustrate the nine possible edge situations, 
also referred to as edge states, in Fig.~\ref{fig:edgestate},
along with the only non-zero transition rates. As in every single spin
flip Markov chain, there is still some freedom in the choice of the transition
probabilities: the detailed balance condition only constrains the ratio
of the transition probabilities between two mutually accessible configurations.
As we aim at finding configurations of large lengths, we set the acceptance
probability to one for allowed transitions increasing the number of present 
edges, which automatically fixes the transition probability of the reversed
move (see Fig.~\ref{fig:edgestate}). This choice is of course possible only
for values of the parameters $u$, $\epsilon$ and $\eta$ such that these
probabilities are smaller than one.

It can be more intuitive to envision the probability 
measure~(\ref{eq:relaxed_prob}) as proportional to $\exp[-E(\uS)]$,
where the ``energy'' $E$ is infinite for forbidden configurations, and
otherwise equal to
\begin{equation}
E_{u,\eta,\epsilon}(\uS)= - \left(N-\sum_l S_l\right) \log u -
\left(n_{\uS}-1\right)\log \eta  -
n_{\epsilon} \log \epsilon \;.
\label{eq:cost_fction}
\end{equation}
In the relevant situation here, $u>1$ and $\epsilon,\eta<1$, a Hamiltonian cycle
of a graph, provided it exists, corresponds to a ground state of this energy
function (of zero energy with the normalizations we chose).

A typical approach for finding low energy configurations in a complex
system is to use a  simulated annealing~\cite{anneal} version of an MC
algorithm (or an implementation of parallel tempering~\cite{pt}).
Starting from a random configuration, one slowly reduces the
value of the parameter conjugate to the energy function 
(typically called the temperature) in order to
obtain a state of minimal energy. 
In our specific case we have a very complex form of the energy
function~(\ref{eq:cost_fction}), and setting up an effective
scheduling for the three relevant external parameters would not be
straightforward. We choose instead to start the MC algorithm with
the initially empty configuration and run it at fixed parameter values.
Long (possibly Hamiltonian) cycles thus appear as fluctuations
around the equilibrium state which is determined by the chosen set of
parameters values. 

The rejection rate of the stochastic process we have defined is very
high: we typically get average acceptance rates of the order of
${\cal O}(1/M)$.  Because of that the implementation of a rejection free
version of the algorithm, inspired by the well-known $N$-fold
algorithm~\cite{nfold}, is of very substantial help.  At each time
step we maintain a list of all possible moves along with their
acceptance probability. Now there is no rejection but the clock is
stochastic. The technical details 
about the necessary bookkeeping are presented in Appendix~\ref{appendix_nfold}.

The combination of the a priori
simplistic approach of working at a fixed value of the external parameters
and the improvement due to the rejection free implementation is
quite efficient in practice, as will become clear in the next section.

\subsection{Numerical Results}

The stationary distribution (cf. Eq.~\ref{eq:relaxed_prob}) reached at
long times by the above described random walk has a positive probability
on the set of the longest cycles of the graph under study.
If the graph is Hamiltonian, at some point the Markov chain will come 
across one of the Hamiltonian cycles,
providing a positive answer to the existence problem and solving at the 
same time the finding one. How
fast such a Hamiltonian configuration is encountered depends on the values
of the parameters defining the transition probabilities. We find that
in the regime of sizes and connectivity distributions we studied, the choice
$u \sim 10^3$, $\epsilon \sim 0.99$ and $\eta \sim 0.1$ led to surprisingly
good results.

Indeed, running the Monte Carlo algorithm (in its rejection free 
implementation) on the same set of graphs that was studied in the previous
section~\footnote{For the largest size $N=1600$, we restricted the sample of
graphs to those on which the decimation strategy had proved unsuccessful.}, we 
find a Hamiltonian cycle in all of them in a reasonable time frame, including 
those for which the decimation strategy was not able to prove their 
Hamiltonianicity.

\begin{figure}[!ht]
  \centering
  \subfigure{\includegraphics[angle=270]{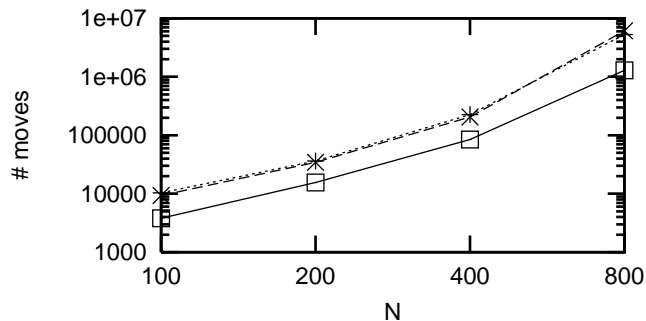}}
  \caption{\label{fig:mediannbmoves_Nfold} 
    Median of the number of moves performed before the discovery of an 
Hamiltonian circuit for various connectivity distributions 
($q_{3,4}^{0.5}$: dotted line (+), $q_{3,5}^{0.5}$: dashed line ($\times$), 
$q_{4,5}^{0.5}$: full line ($\Box$)) in function of the size of the graphs.
}
\end{figure}

It is however expected that the external parameters should be tuned with
the size of the graphs. We plot
in Fig.~\ref{fig:mediannbmoves_Nfold} the (median) number of moves 
performed before the discovery of a Hamiltonian circuit for graphs drawn 
from the three connectivity distribution ensembles of various sizes. This
grows exponentially with the sizes of the graphs, consistent with the 
picture that Hamiltonian cycles are found as a fluctuation
of the observables $L$, $n_{\uS}$ and $n_\epsilon$ from their typical 
values.

For comparison with the decimation strategy, we mention 
it takes about $40$ minutes  to find a Hamiltonian
cycle using the MC algorithm with reasonably optimized parameter
values for the same graph of size $N=1600$ and degree distribution
$q_{3,4}^{0.5}$ of the previous section. Of course, this rather good
result does not include the time it takes to find these optimized
values for the three parameters $u$, $\epsilon$ and $\eta$.
 
\section{Conclusions}
\label{sec_conclusions}

We have introduced two distinct methods to locate long cycles
in graphs and tested them on random graphs of minimal 
degree 3. For all of the many instances we investigated, at least one
of the two methods was able to construct a Hamiltonian cycle. After the
non-rigorous statistical mechanics study of~\cite{MaSe}, this constitutes
another constructive confirmation of the conjecture put forward 
in~\cite{Wormald}, according to which these graphs are, with high 
probability, Hamiltonian.

The two algorithms presented in this paper are of a very different nature, it
is thus difficult to assert their relative efficiency. The belief inspired
decimation procedure of Sec.~\ref{sec_decimation} has the advantage of
versatility: there are few parameters and their influence is not crucial.
Moreover its computational cost on this family of sparse random graphs seems
to grow polynomially with the number of vertices. Its caveat is that it
does not always construct a Hamiltonian cycle (even if constructing a cycle
of extensive length can already be seen as a positive result) of the random
graphs on which we tested it. This motivated the development of the Monte
Carlo approach of Sec.~\ref{sec_mc}, which turned out to be successful on
every investigated graph. This method relies on a fortunate trade-off between
the simplicity of the underlying idea and the efficiency of the elaborate
(rejection free) implementation this simplicity allows. At variance with
the decimation procedure, this method is highly sensitive to its parameters,
that have to be carefully determined by trial and error. 

We see as a possible continuation of this work a more systematic investigation
of the Monte Carlo method, in particular on the automatic adjustment of 
the optimal parameters with the size of the graphs and the connectivity
distribution. In this respect Tempering and Parallel tempering~\cite{pt} 
could constitute useful approaches to this issue.

Another possible direction for future work would be to apply these algorithms
to real-world networks~\cite{review_networks}. In this context an 
interesting issue would be the study of intermediate length circuits, 
i.e.~those which are too long to be found by exhaustive enumeration,
yet much shorter than the total size of the graphs (this intermediate scale
seems the most relevant to discuss for example routing in Internet networks). 
The two methods presented here can be easily adapted to tackle this problem.

Finally, we mention that the largest graph on which we found a
Hamiltonian cycle in a reasonable CPU time (30 minutes, by using the decimation
strategy combined with the local rewiring) was a
random mixture graph of size $N=12800$ and degree distribution
$q_{3,4}^{0.5}$.



\acknowledgments
This work was supported by EVERGROW, integrated project No.~1935 in the
complex systems initiative of the Future and Emerging Technologies
directorate of the IST Priority, EU Sixth Framework.

\appendix

\section{Patching Vertex Disjoint Cycles by Local Rewiring}
\label{app:LR}
 
Here we describe the ``patching'' procedure we apply to cycle covers
made of several vertex disjoint cycles obtained at the end of a
decimation procedure.  It aims at uniting these distinct cycles into a
single Hamiltonian one.

\begin{figure}[!ht]
  \centering
  \includegraphics[width=0.3 \textwidth]{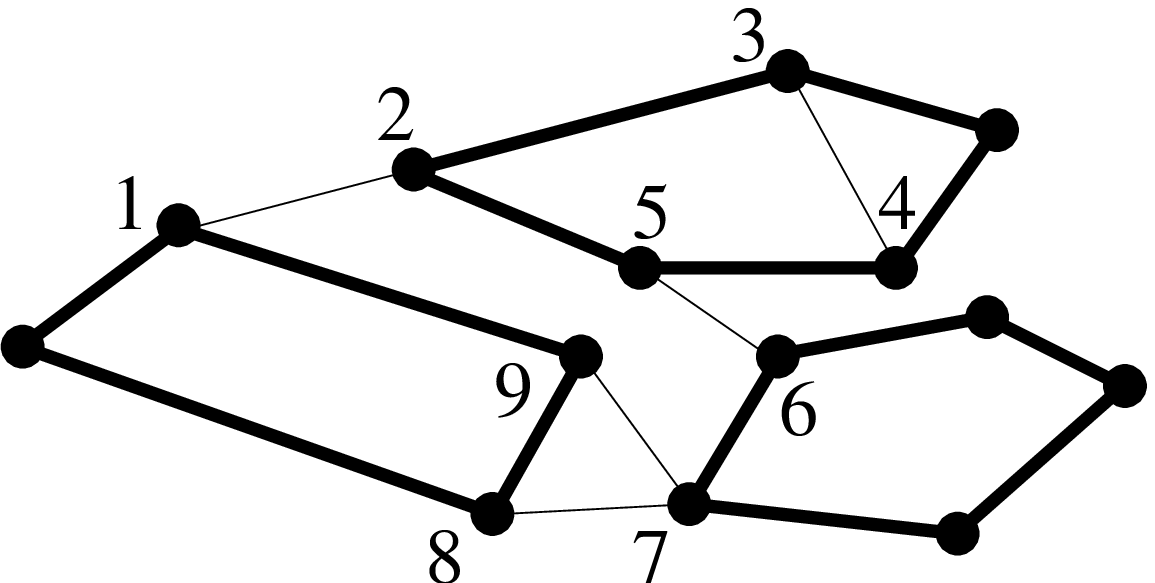}
  \hspace{2cm}
  \includegraphics[width=0.3 \textwidth]{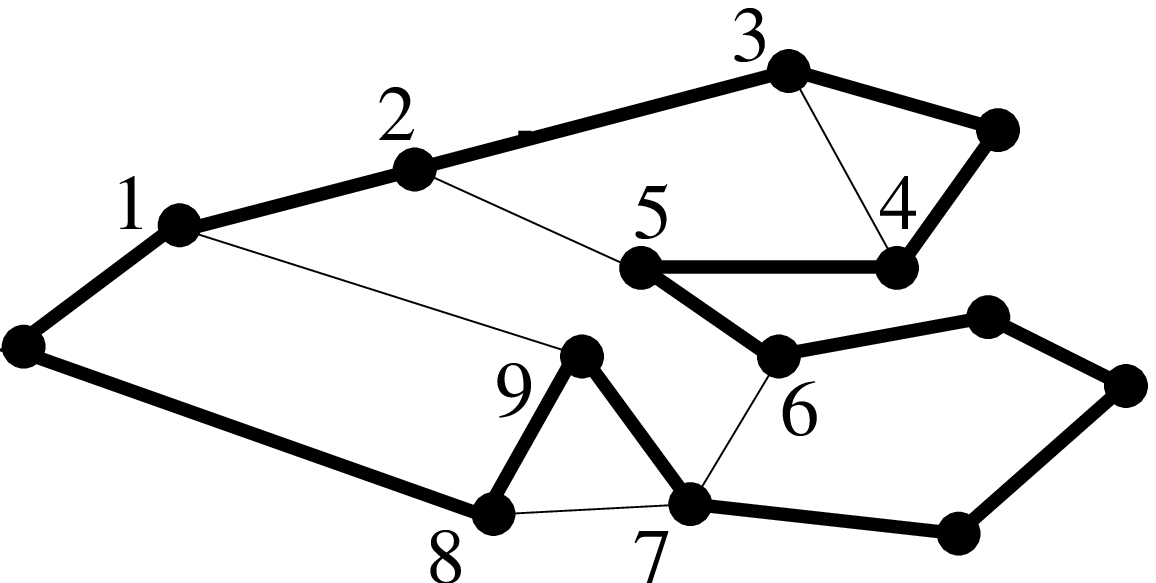}
  \caption{\label{fig:lm} Local rewiring in order to unite the cycles of a cycle cover (left-hand side) into a Hamiltonian cycle (right-hand side). }
\end{figure}

As mentioned in section~\ref{sec:success}, the cycle covers we find
are typically made of a long (extensive) cycle, and a few small
ones. This suggests that in order to unite them we should look for
some simple displacement of the edges around the small cycles in order
to connect them among themselves and with the longest one.  The
patching procedure we adopt consists in removing two or more edges
belonging to different cycles of the considered cycle cover.  At
the same time, we introduce an equal amount of edges which were not
present in the original cycle cover subgraph (but are part of the graph), which
close the gaps we created in the vertex disjoint cycles and unite
them into one cycle.  For example, given the cycle cover presented on
the left of Fig.~\ref{fig:lm} we could transform it into the
Hamiltonian cycle presented to its right by changing the edge
variables of the edges $\{1,2\}, \{2,5\}, \{5,6\}, \{6,7\}, \{7,9\}$
and $\{1,9\}$ to their complementary value. We will refer to this
process as \textit{local rewiring}.

The problem now is to determine a strategy
for finding this appropriate set of
edges which allows us to change a cycle cover into a Hamiltonian cycle
after having performed this local reconnection. We explain our rules
on the example drawn in Fig.~\ref{fig:lm}.  

We start from some arbitrary
vertex belonging to the smallest cycle of the cycle cover, say~$1$.  
If vertex $1$
has no neighboring edge of which the other ending vertex lies on a
different cycle of the cycle cover we are considering, we move to the
next vertex on the cycle we started from (e.g.~$9$). In case however
it does (as is the case in Fig.~\ref{fig:lm} due to the presence of
edge $\{1,2\}$), we check whether the same holds for one of the
neighboring edges ($3$ or $5$) of the vertex on the new cycle
($2$). If this edge leads to the original cycle we started from (which
is clearly not the case for edge $\{3,4\}$, but would be for $\{5,9\}$
or $\{5,8\}$ if they were present), and if it
ends in a neighboring vertex of the original vertex we started from
($1$), we have found a valid update:
it consists in changing the values of all edge variables included in the path
going from the first to this last vertex. 
For example, if the edge $\{5,9\}$ were part of the graph we
would have found a subset of edges (i.e.~$\{1,2\}, \{2,5\}, \{5,9\}$
and $\{1,9\}$) for which, if we set their corresponding edge
variables to their complementary value, we would have united two of
the three cycles. In such a case, i.e.~when after some local rewiring
we have reduced the number of cycles in the cycle cover, but not yet
to just one, we go to the next vertex on the now extended cycle we
started from (which would then be vertex $3$), and follow the same
rules in order to unite it to the still remaining vertex-disjoint
cycles.  This is not the case in the figure: edge
$\{5,6\}$ brings us yet to another cycle and
again we look for a way to reach a different cycle through the present,
neighboring edges of this new vertex ($6$). We continue this procedure
until we get back to the original cycle, and make sure we never cross
the same cycle twice, as it would not lead to a unique Hamiltonian
cycle.

We do this operation for all the vertices belonging to the smallest
cycle (which throughout the local rewiring could grow in length). The
number of operations we need to perform for each one of these vertices
depends on its degree, which influences the number of edges to other
cycles we need to consider, and on the number of cycles composing the cycle
cover. As we only considered graphs with a maximum degree up
to $5$, the number of edges of a vertex leading to another cycle
is not larger than $3$.  Due to the fact that cycle covers typically contain
one large cycle, the total number of cycles composing it is usually
quite low (it never exceeded $14$ for all the graphs we
investigated). Hence the local rewiring requires a number of operations
negligible compared to the cost of the decimation procedure.

Even though it 
turns out to be quite effective in practice,
we must stress that the patching
procedure we presented here, i.e.~the local rewiring of the edges,
is rather restrictive. Its only goal is to immediately try to reduce
the number of cycles composing the cycle cover, not to sample the set of all
cycle covers.

\section{Details on the Implementation of the Rejection Free Algorithm}
\label{appendix_nfold}

As mentioned in section~\ref{sec_mc}, an MC sweep in the MC
algorithm with rejection consists in proposing $M$ changes among the
$M$ edges.  The probability with which a move is accepted depends on
the probability law~(\ref{eq:relaxed_prob}) according to the detailed
balance condition. We have explicitly given 
the non-zero transition rates in Fig.~\ref{fig:edgestate}.
It turns out that the actual
number of accepted moves during a sweep is only of order ${\cal O}(1)$ for all 
graphs we investigated.

An $N$-fold \cite{nfold}, rejection free version of the Monte Carlo
simulation helps to alleviate this problem.  Rather than proposing a
change which is then possibly discarded, we choose a change of non-zero
probability and compute the (random) number of rejections that would have
occurred before its acceptation. To this aim we maintain lists of the edges 
in a given state $j$ (see Fig.~\ref{fig:edgestate}) and the sizes $n_j$ of 
these lists. 
The escape probability from the current configuration is then simply given by 
\begin{equation}
P_{esc}=\frac{1}{M} \sum_{\uS'}W(\uS \rightarrow
\uS')= \frac{1}{M} \sum_{j=1}^9 n_j w_j,
\label{eq:escape_probability}
\end{equation}
where we denote $w_j$ (instead of $w_{j \to i}$ in Fig.~\ref{fig:edgestate}) 
the probability with which an edge in state $j$ changes status.

Each proposed (and necessarily accepted) move of the rejection free MC
algorithm now consists in drawing an edge state $i$ with probability
$n_i w_i / (M P_{esc})$. The edge that will be changed is uniformly
drawn from the $i$'th list and the clock is increased by an amount
$-\log r / (M P_{esc})$, where $r$ is a uniform random variable drawn
in $[0,1]$.

The bookkeeping of the state of the edges is negligible in terms of 
space requirements. There is however an overhead in the number of operations
performed at each step: the modification of edge $l$ changes the state of 
other edges. Most of the time these are only the direct neighbors of 
$l$, which are few for the low degree graphs we investigate. 
However, when breaking up a cycle into a path or vice versa 
($1 \leftrightarrow 5$), all edges belonging to the cycle have to be
updated. Note that even in the very unrealistic case where long cycles 
(of order $N$) are broken or created at each step, this overhead is still
compensated by the gain with respect to the usual MC algorithm where the
acceptance rate is of order $1/M$.

\begin{figure}[!ht]
  \centering
  \subfigure{\includegraphics[angle=270]{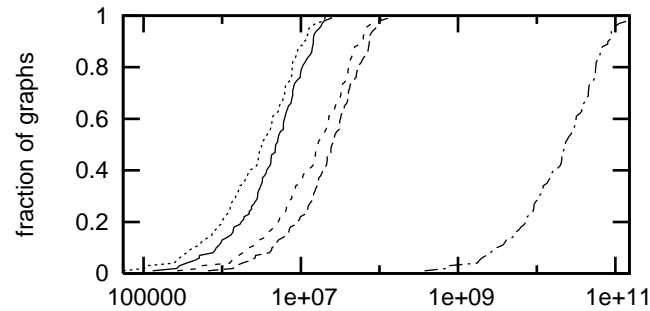}\vspace{0.11cm}}
    \caption{\label{fig:soltime_nbmoves} Distribution of the
  solution time, i.e.~the time it takes to find a Hamiltonian cycle, in
  stochastic MC time for $N$-fold MC (long-dashed line) and MC sweeps
  for rejection MC (short-dashed line), of the number of moves
  performed (full line for $N$-fold MC, dotted line for rejection MC)
  and of the total number of proposed moves in rejection MC
  (dashed-dotted line) for graphs of size $N=800$ with degree
  distribution $q_{3,4}^{0.5}$. }
\end{figure}

Figure~\ref{fig:soltime_nbmoves} contains some numerical evidence that
the $N$-fold MC is indeed faster than the rejection MC.  The distribution of 
the time (number of sweeps for the simple MC or stochastic clock for the 
rejection free version) at which a Hamiltonian cycle is found is almost the same for 
both algorithms, as it should be. The same conclusion holds for the number of 
moves actually performed before this event. However, the number of proposed
moves in the rejection MC is higher by a large factor, inversely proportional to the acceptance rate.

The computational overhead due to the update of the edge states does
not spoil this gain: it takes $43$ seconds for the rejection free
implementation to find a Hamiltonian cycle in an exemplary graph of
size $N=800$ and degree distribution $q_{3,4}^{0.5}$, while the
rejection MC requires more than $10$ hours.

\end{document}